\documentclass[showpacs,preprintnumbers,amsmath,amssymb,prb,
twocolumn,superscriptaddress]{revtex4}

\usepackage{graphicx}   % Include figure files
\usepackage{dcolumn}    % Align table columns on decimal point
\usepackage{bm}         % bold math

\begin{document}

\title{Spin detection at elevated temperatures using a driven double quantum dot}

\author{G. Giavaras}
\affiliation{Department of Materials, University of Oxford, Oxford
OX1 3PH, UK}

\author{J. Wabnig}
\affiliation{Department of Materials, University of Oxford, Oxford
OX1 3PH, UK}\affiliation{Cavendish Laboratory, Department of
Physics, University of Cambridge, Cambridge CB3 0HE, UK}

\author{B. W. Lovett}
\affiliation{Department of Materials, University of Oxford, Oxford
OX1 3PH, UK}

\author{J. H. Jefferson}
\affiliation{QinetiQ, St. Andrews Road, Malvern WR14 3PS, UK}

\author{G. A. D. Briggs}
\affiliation{Department of Materials, University of Oxford, Oxford
OX1 3PH, UK}

%\date{\today}


\begin{abstract}
We consider a double quantum dot in the Pauli blockade regime
interacting with a nearby single spin. We show that under
microwave irradiation the average electron occupations of the dots
exhibit resonances that are sensitive to the state of the nearby
spin. The system thus acts as a spin meter for the nearby spin. We
investigate the conditions for a non-demolition read-out of the
spin and find that the meter works at temperatures comparable to
the dot charging energy and sensitivity is mainly limited by the
intradot spin relaxation.
\end{abstract}
\maketitle

\section{Introduction}

Electron spins in semiconductors and molecular systems are good
candidates for qubits due to their relatively long coherence
times. Manipulation of single spins and controlled interaction
between pairs of spins are essential ingredients for quantum
information processing. Single spin rotation has been demonstrated
in electrostatically defined quantum dots using the electron spin
resonance technique.\cite{koppens1} Coherent manipulation of a
pair of qubits, giving rise to entanglement, has also been
achieved in a semiconductor double dot (DD) device based on fast
electrical pulses and operating the dots in the spin blockade
regime.\cite{petta} Single spin rotations together with
entanglement generation in principle enable universal quantum
operations. A further operation for a quantum processor is spin
detection, which is essential for projection of the quantum state
after computation and read-out of the result. This is the main
focus of this paper.

Single spin detection is also important for future spintronic
devices in general, and various electrical and optical schemes
have been proposed and demonstrated. For example Elzerman $et$
$al.$ demonstrated experimentally a single-shot read-out of a
quantum dot spin using a spin to charge conversion
technique,\cite{elzerman} while Rugar $et$ $al.$ employed magnetic
resonance force microscopy to probe the state of a single
spin.\cite{rugar} Other schemes for spin read-out involve open
quantum dots with an inhomogeneous Zeeman splitting and closed DD
systems which are coupled to quantum point
contacts.\cite{barrett,engel} It has been shown theoretically that
the dc-electrical current and shot noise through the dots or the
point contacts can provide valuable information about the spin
state, the energy spectrum and the relevant decoherence
rates.\cite{barrett,engel} However, for optically non-active
molecular spins no reliable read-out scheme exists.

\begin{figure}
\begin{centering}
\includegraphics[height=4.2cm]{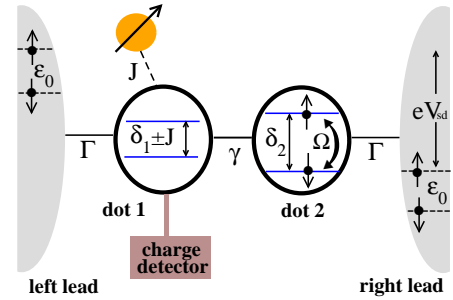}
\par\end{centering}
\caption{Schematic illustration of the proposed system for a
single spin detection. Two tunnel-coupled quantum dots are
connected to leads enabling current to pass through. A nearby spin
interacts with the spins on dot 1 and this interaction induces a
different Zeeman splitting in the two dots. In an external
magnetic field and under microwave irradiation the spins on the
two dots can be rotated with a Rabi frequency $\Omega$ and the
average electron occupation exhibits resonances which are
sensitive to the state of the nearby spin. A charge detector is
capacitively coupled to dot 1 and is used to monitor a change in
the occupation. In the rotating frame with respect to the incident
microwave field the lead electrons have a Fermi energy that
depends on their spin state with a relative energy difference
$\varepsilon_{0}=\hbar\omega_{0}$ where $\omega_{0}$ is the
frequency of the driving magnetic field. $V_{sd}$ is the bias
voltage across the dots. The effective Zeeman splitting in each
dot, $\delta_{1}\pm J$ and $\delta_{2}$ respectively depends on
both the orientation of the nearby spin and the interaction
strength $J$.} \label{exchange}
\end{figure}

In this work we consider a single spin (target spin) that
interacts with the spins of two tunnel-coupled quantum dots, as
shown schematically in Fig.\ref{exchange}, and demonstrate how to
probe its state by monitoring the average electron occupation on
one of the two dots. The interaction between the target spin and
the spins on the DD induces an effective Zeeman splitting that is
different in each dot. Also, the sign of the Zeeman splitting
depends on the orientation of the target spin. This target-spin
dependent asymmetry of the Zeeman splitting makes it possible to
rotate only one of dot spins and thus results in a target-spin
dependent lifting of the Pauli blockade. We show this lifting of
the blockade as a change in the average dot occupation, which can
be measured by a charge detector. Alternatively, the change in
occupation is directly related to a change in the current through
the double dot, which, however, might be too small to detect by
standard dc-measurement techniques.

In order to realize a non-demolition spin measurement the spins on
the DD and the target spin must have different Zeeman splitting,
most likely to be achieved by different $g$-factors through
$g$-factor engineering or choice of materials. A general spin-spin
interaction always contains so called spin flip terms, which are
suppressed when the difference in the Zeeman splitting of the dot
electrons and the target spin is larger than the spin-spin
interaction strength.

A theoretical investigation has shown that a single quantum dot
works as a spin meter, but the proposed device has the drawback of
operating only at low temperatures, comparable to the energy scale
set by the Zeeman spitting.\cite{wabnig1} In this article we show
that the driven DD spin detector works at much higher
temperatures, comparable to the dot charging energy and we find
that the main limitation to its sensitivity is intradot spin
relaxation. Thus we present a scheme that allows the spin state of
a nearby spin to be probed noninvasively in a single shot, using
different electron occupations on the double dot for spin up and
spin down orientations as a basic readout mechanism.

In the following section we introduce our model and discuss
technical details of the solution. In Sec.~\ref{sec:Results} we
show how the double dot acts as a spin meter and explore how
interdot hopping, microwave intensity, temperature, spin-spin
interaction strength and spin relaxation influence its
performance. We conclude in Sec.~\ref{sec:Conclusions} by
discussing the operation of the spin meter for experimentally
accessible parameters.

\section{Physical Model}

The total system consists of the DD, a nearby spin, metallic leads
and a bosonic heat bath. This system is modelled by the
Hamiltonian
\begin{equation}
H_{tot}=H_{S}+H_{leads}+H_{T}+H_{B}+H_{SB},\notag
\end{equation}
where $H_{S}$ models the DD and the spin, $H_{leads}$($H_{B}$)
models the leads (heat bath) and $H_{T}$($H_{SB}$) models the
interaction between the leads (heat bath) and the DD.
Specifically, for the DD and the nearby target spin we write the
Hamiltonian as
\begin{equation}
H_{S}=H_{DD}+H_{M}+H_{I},\label{system}
\end{equation}
where $H_{DD}$ is a Hubbard Hamiltonian describing the DD, $H_{M}$
is due to the applied magnetic fields and $H_{I}$ models the
interaction of the nearby spin with the DD system. For the DD we
have
\begin{equation}
\begin{split}H_{DD}= & \sum_{i=1}^{2}\varepsilon_{i}n_{i}-\gamma\sum_{\sigma}(c_{1\sigma}^{\dagger}c_{2\sigma}+c_{2\sigma}^{\dagger}c_{1\sigma})\\
& +U\sum_{i=1}^{2}n_{i\uparrow}n_{i\downarrow}+Vn_{1}n_{2},
\end{split}\label{ddot}
\end{equation}
that allows up to 2 electrons per dot. The number operator is
$n_{i}=\sum_{\sigma}n_{i\sigma}=\sum_{\sigma}c_{i\sigma}^{\dagger}c_{i\sigma}$
for dot $i=\{1,2\}$ and spin $\sigma=\{\uparrow,\downarrow\}$. The
operator $c_{i\sigma}^{\dagger}$ ($c_{i\sigma}$) creates
(annihilates) an electron on dot $i$ with on-site energy
$\varepsilon_{i}$. $\gamma$ is the tunnel coupling between the two
dots, $U$ is the charging energy (intradot Coulomb energy) and $V$
the interdot Coulomb energy. The Hamiltonian part due to the
applied magnetic fields, that breaks the spin degeneracy, is
\begin{equation}
\begin{split}
H_{M}= & \sum_{i=0}^{2}\frac{\Delta_{i}}{2}\sigma_{i}^{z}+\sum_{i=1}^{2}\hbar\Omega\cos(\omega_{0}t)\sigma_{i}^{x},\label{bfield}
\end{split}
\end{equation}
where $i=0$ refers to the target spin and the spin operators are
defined in the standard way
$\bm{\sigma}_{i}=\sum_{\sigma\sigma'}c_{i\sigma}^{\dagger}\bm{\sigma}_{\sigma\sigma'}c_{i\sigma'}$,
with $\bm{\sigma}$ being the vector of the $2\times2$ Pauli
matrices. $\Delta_{i}=g_{i}\mu_{B}B_{i}$ is the Zeeman splitting
due to a static magnetic field $B_{i}$ along $z$, and a $g$-factor
$g_{i}$, $\Omega$ is the Rabi frequency and $\omega_{0}$ the
frequency of the oscillating magnetic field along $x$. For a
single spin the oscillating magnetic field rotates the
$z$-component of the spin with frequency $\Omega$ when
$\Delta=\hbar\omega_{0}$. We have ignored the effect of the
oscillating field on the target spin which is a good approximation
for narrow-band radiation that is only resonant with the spins on
the DD (or alternatively only with the target spin), a condition
that can be achieved for example by engineering different
$g$-factors in the dots and the target spin.

Moreover, we assume that the target spin interacts only with dot
1, although the basic idea can be extended to the most general
case when the target spin interacts with both dots. As shown below
our scheme is still efficient provided that the strength of the
interaction between the target spin and each dot is different, a
condition that is typically satisfied. We consider an Ising
interaction between dot $i=1$ and the target spin $i=0$ of the
form
\begin{equation}
H_{I}=\frac{J}{2}\sigma_{0}^{z}\sigma_{1}^{z},\label{inter}
\end{equation}
with $J$ being the strength of the interaction. $J$ mainly depends
on the distance of dot 1 from the target spin as well as the
actual size of the dot and the target spin. Physical values for
$J$ for a purely dipolar interaction are within the range of a few
MHz as shown in Ref. 7. This form of interaction is justified when
there is negligible tunnel-coupling between dot 1 and the nearby
spin so that to a good approximation hopping can be ignored. In
addition, spin-flip processes are weak due to the Zeeman splitting
induced by the static magnetic field and thus neglected. This is a
good approximation when the difference in the Zeeman splittings
between the target spin and the DD spins is much larger than the
interaction strength $J$. Under these conditions the Ising
interaction Eq.(\ref{inter}) is a reasonable choice and leads to a
non-demolition measurement.

The choice of Ising interaction dictates that the combination of
quantum dot system and target electron has to be specifically
tailored to realise this non-demolition measurement. The necessary
regime of parameters might be difficult to realise in a
gate-defined quantum dot system, e.g. in GaAs, but arises quite
naturally in carbon nanotube dots probing a molecular spin. For
example with a typical dipole-dipole spin interaction strength of
5 MHz, a typical difference in Zeeman splitting of about 5\% and a
typical EPR Zeeman splitting of 10 MHz (see also Ref.~7) we arrive
at a ratio of coupling strength to difference in Zeeman splitting
of 1/100. Considering the flip flop terms as a perturbation to the
diagonal Hamiltonian as in Ref.~8, the first order corrections
vanish and only second order terms contribute, which are typically
suppressed by a factor of 1/10000. This means that any spin flip
that can disturb the measurement will take place at a much reduced
rate, thus making a non-demolition measurement possible.

For the DD with single orbital levels there are in total 16 states
and the maximum number of electrons is 4. The many-body states of
the DD and target spin system can be written in the form
$|DD\rangle|\sigma_{T}\rangle$ where $|DD\rangle$ denotes the
many-body states of the DD and
$|\sigma_{T}\rangle=|\uparrow\rangle$ or $|\downarrow\rangle$
denotes the two possible target spin states. Starting with
Eq.(\ref{system}) we can define the two uncoupled spin up,
$H_{S}^{+1}$, and spin down, $H_{S}^{-1}$, Hamiltonians in the
subspace $|DD\rangle|\uparrow\rangle$ and
$|DD\rangle|\downarrow\rangle$ respectively depending on the state
of the target spin. From Eqs.(\ref{ddot}),(\ref{bfield}) and
(\ref{inter}) it can be shown that
\begin{equation}
\begin{split}
H_{S}^{\sigma}= &H_{DD}+\sigma\frac{\Delta_{0}}{2}+\frac{\Delta_{1}+\sigma J}{2}\sigma_{1}^{z}+\frac{\Delta_{2}}{2}\sigma_{2}^{z}\\
&+\sum_{i=1}^{2}\hbar\Omega\cos(\omega_{0}t)\sigma_{i}^{x},\label{hupdown}
\end{split}
\end{equation}
with $\sigma=+1(-1)$ for target spin up(down).

\begin{figure}
\begin{centering}
\includegraphics[height=5.5cm]{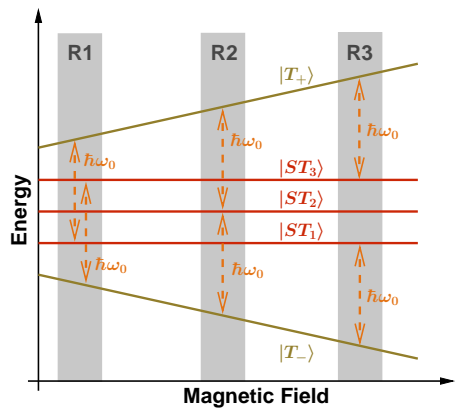}
\par\end{centering}
\caption{Schematic illustration of the two-electron energy diagram
for the undriven system. $\omega_{0}$ is the oscillating frequency
of the driving field. $\left|T_{+}\right\rangle $and
$\left|T_{+}\right\rangle $ are blocked triplet states,
$\left|ST_{i}\right\rangle ,\,i=1,2,3$ are states with a singlet
component. R1, R2 and R3 label regions of possible resonances}
\label{diagram}
\end{figure}

Figure~\ref{diagram} shows schematically the two-electron energy
level diagram as function of the static magnetic field\cite{note1}
of the undriven system ($\Omega=0$) and for spin up. The upper and
lower energy levels correspond to the triplets $|T_{+}\rangle$ and
$|T_{-}\rangle$ which as explained in the next section lead to
spin blockade. The remaining three energy levels,
$|ST_{i}\rangle$, $i=1,2,3$, correspond to two-electron states
that contain a singlet component and thus allow electronic
transport. For the driven system ($\Omega\neq0$) transitions from
the $|T_{\pm}\rangle$ to $|ST_{i}\rangle$ states can take place
lifting the spin blockade and resulting in a change of the average
occupation. Within a simplified approach if a transition frequency
matches the frequency of the driving field the corresponding
transition probability is expected to be high. The regions R1, R2,
R3 in Fig.~\ref{diagram} illustrate such a resonant behaviour. The
magnetic field where the resonances occur depends on the energy
splitting of the $|ST_{i}\rangle$ states and therefore on the
interdot hopping $\gamma$ and the coupling strength $J$.
Additionally, as shown in the next section, interference effects,
for instance when $|T_{\pm}\rangle\rightarrow |ST_{2}\rangle$, can
yield a vanishingly small probability for particular magnetic
fields.

Having described the interaction of the DD with the oscillating
magnetic field we analyse the dissipative interactions with the
leads and the phonon heat bath. The left and right leads are
described by a Hamiltonian of the form $H_{leads}=\sum_{\ell
k\sigma}\epsilon_{\ell k}d_{\ell k\sigma}^{\dagger}d_{\ell
k\sigma}$, where $d_{\ell k\sigma}^{\dagger}$ ($d_{\ell k\sigma}$)
creates (destroys) an electron in lead $\ell=\{L,R\}$ with
momentum $k$, spin $\sigma$ and energy $\epsilon_{\ell k}$. The
interaction between the dots and the leads is given by the
tunneling Hamiltonian
\begin{equation}
H_{T}=\sum_{k\sigma}(t_{L}c_{1\sigma}^{\dagger}d_{Lk\sigma}+t_{R}c_{2\sigma}^{\dagger}
d_{Rk\sigma})+\text{H.c.},
\end{equation}
where $t_{L}(t_{R})$ is the tunnel coupling between dot $1(2)$ and
lead $L(R)$ and we consider the symmetric case where
$t_{L}=t_{R}$.

To take into account spin relaxation we have considered a generic
bosonic bath that is modelled as a set of harmonic oscillators and
is described by the Hamiltonian
$H_{B}=\sum_{j}\hbar\omega_{1,j}a_{1,j}^{\dagger}a_{1,j}+\sum_{j}\hbar\omega_{2,j}a_{2,j}^{\dagger}a_{2,j}$.
We have assumed that each quantum dot is coupled to an independent
bosonic bath and there are no environment-induced correlations
between the two dots. The operators $a_{1,j}^{\dagger}$
($a_{1,j}$) create (destroy) a boson in mode $j$ and similarly for
$a_{2,j}^{\dagger}$ ($a_{2,j}$), while $\omega_{1,j}$ are the
corresponding frequencies of the bath modes. The interaction
between the bath and the spins of the DD is given by the general
model Hamiltonian
\begin{equation}
H_{SB}=\sigma_{1}^{-}\sum_{j}\Lambda_{1,j}a_{1,j}^{\dagger}+\sigma_{2}^{-}\sum_{j}\Lambda_{2,j}a_{2,j}^{\dagger}+\text{H.c.},
\label{bathint}
\end{equation}
where the spin-flip operators are
$\sigma_{i}^{-}=c_{i\downarrow}^{\dagger}c_{i\uparrow}$ and
$\Lambda_{1,j}(\Lambda_{2,j})$ is the coupling constant between
dot $1(2)$ and the $j$th mode of the corresponding bath. $H_{SB}$
allows spin-flip processes for electrons in the DD via energy
exchange with the bath which, as shown in the next section, leads
to a leakage current. We consider spin relaxation only in the DD
since for the target spin an upper limit to its relaxation rate is
set implicitly by the coupling to the leads. In our scheme the
relaxation rate of the target spin has to be smaller than the
electron tunneling rate from the leads to the DD to ensure a
measurable change in the DD occupations before spin relaxation.

To investigate the electron occupation of the system we employ a
master equation approach\cite{gardiner} and derive an equation of
motion for the reduced density matrix, $\rho$, for the system of
interest that consists of the DD and the target spin. The
occupation probabilities are given by the diagonal elements of
$\rho$. It is convenient to eliminate the time dependence from the
system Hamiltonian $H_{S}^{\sigma}$ and for this reason we perform
a rotating wave approximation.\cite{gardiner,wabnig2} This
approximation is well-justified only for weak driving, i.e., when
$\Omega\ll\omega_{0}$ as in our system. In the rotating frame an
arbitrary system operator $\mathcal{K}$ is transformed as
$U_{z}^{\dagger}\mathcal{K}U_{z}$ with the unitary operator
$U_{z}=\exp(-i\sigma^{z}\omega_{0}t/2)$ and
$\sigma^{z}=\sum_{i=0}^{2}\sigma_{i}^{z}$.

Starting with the total density matrix, $\chi$, and within the
standard Born and Markov approximations\cite{gardiner} we derive
an equation of motion for $\rho$ by tracing over the leads and
bosonic bath degrees of freedom, i.e., $\rho$=Tr$_{E}\{\chi\}$
where $\text{Tr}_{E}\{...\}$ means trace over the environmental
degrees of freedom. In the rotating frame and having performed a
rotating wave approximation the density matrix $\rho$ satisfies
the equation of motion
\begin{equation}
\dot{\rho}(t)=\mathcal{L}_{S}\rho(t)+\mathcal{L}_{leads}\rho(t)+\mathcal{L}_{B}\rho(t),\label{master}
\end{equation}
with the free evolution term
\begin{equation}
\mathcal{L}_{S}\rho(t)=-\frac{i}{\hbar}[\mathcal{H}_{S}^{\sigma},\rho(t)]\notag,
\end{equation}
and the terms due to the electronic leads
\begin{equation}
\begin{split}
\mathcal{L}_{leads}\rho(t)= & -\frac{1}{\hbar^{2}}\text{Tr}_{E}\{\int_{0}^{\infty}d\tau[H_{T}(t),\\
&[U(\tau)H_{T}(t-\tau)U^{\dagger}(\tau),\rho(t)\otimes\rho_{leads}]]\}\notag,
\end{split}
\end{equation}
and the bosonic bath
\begin{equation}
\begin{split}
\mathcal{L}_{B}\rho(t)= & -\frac{1}{\hbar^{2}}\text{Tr}_{E}\{\int_{0}^{\infty}d\tau[H_{SB}(t),\\
&[V(\tau)H_{SB}(t-\tau)V^{\dagger}(\tau),\rho(t)\otimes\rho_{B}]]\}.\notag
\end{split}
\end{equation}
The operators are
$U(\tau)=\exp[-i(\mathcal{H}^{\sigma}_{S}+H_{leads})\tau/\hbar]$
and $V(\tau)=\exp[-i(\mathcal{H}^{\sigma}_{S}+H_{B})\tau/\hbar]$,
with $\rho_{leads}$, $\rho_{B}$ being the equilibrium density
matrix for the leads and the bosonic bath respectively. The time
dependent operators are
$c_{i\uparrow}(t)=c_{i\uparrow}\exp(-i\omega_{0}t/2)$,
$c_{i\downarrow}(t)=c_{i\downarrow}\exp(+i\omega_{0}t/2)$ and the
Hamiltonian $\mathcal{H}^{\sigma}_{S}$ depends on the nearby spin,
$\sigma=+1(-1)$ for spin up(down), i.e.,
\begin{equation}
\mathcal{H}_{S}^{\sigma}=H_{DD}+\sigma\frac{\delta_{0}}{2}+\frac{\delta_{1}+\sigma
J}{2}\sigma_{1}^{z}+\frac{\delta_{2}}{2}\sigma_{2}^{z}+\sum_{i=1}^{2}\frac{\hbar\Omega}{2}\sigma_{i}^{x},\label{hamiltf}
\end{equation}
where we have introduced the magnetic field detuning
$\delta_{i}=\Delta_{i}-\hbar\omega_{0}$.

For the numerical calculations we write Eq.(\ref{master}) in the
energy basis. This results in a system of 256 coupled equations
for all the matrix elements of $\rho$ which is solved numerically
taking into account the normalisation condition for the diagonal
elements, $\sum^{16}_{i=1}\rho_{i,i}=1$. We are interested in the
steady state, $\rho_{st}$, that corresponds to $\dot{\rho}=0$ in
Eq.(\ref{master}). The quantity of interest is the average
electron occupation of the DD, for example of dot 1, that is
calculated as $N_{1}=\text{Tr}\{n_{1}\rho_{st}\}$. In the next
section we present the basic results and explain the influence of
various system parameters on the average electron occupation of
the DD.

\section{\label{sec:Results}Results and discussion}

Before we examine the influence of microwave radiation we have to
make a choice for the operating regime of the DD. DD systems and
their physical response are highly tunable by adjusting the gate
voltages and the source-drain bias voltage in the leads. A regime
which is easily accessible and has attracted a lot of interest is
the Pauli spin blockade regime which has been demonstrated
experimentally in various systems such as AlGaAs/GaAs and Si/Ge
double quantum dots\cite{ono,shaji} as well as carbon nanotube
dots.\cite{buitelaar} In this regime one electron is confined in
each dot and the three triplet states are almost equally and fully
populated. In the absence of spin relaxation and microwaves the
$(1,1)$ triplet state is blocked from moving on to a $(0,2)$ state
by the Pauli exclusion principle ($(n,m)$ denotes a charge state
with $n(m)$ electrons on dot 1(2)). Thus the electrical current as
a function of the source-drain bias is suppressed.

For a fixed source-drain bias in the spin blockade regime a change
in the occupations of the two dots can occur and a
microwave-induced current can flow provided that the two dots have
a different Zeeman splitting.\cite{sanchez} In this case the
oscillating magnetic field in combination with a static field
induces coherent spin rotations that mix two-electron states and
current flows through the transport cycle
$(0,1)\rightarrow(1,1)\rightarrow(0,2)\rightarrow(0,1)$. When the
two dots have the same Zeeman splitting the spins in the two dots
rotate at the same rate in the triplet subspace and therefore the
average occupation remains fixed and current does not
flow.\cite{sanchez,koppens2} In this case only spin relaxation can
give rise to a change in the occupation of the dots.

\begin{figure}
\begin{centering}
\includegraphics[height=6.5cm]{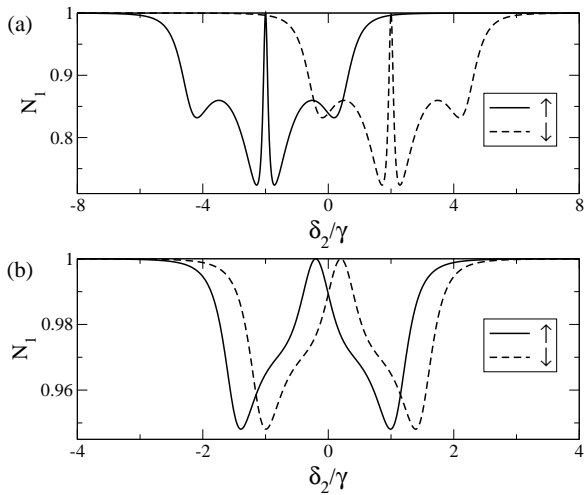}
\par\end{centering}
\caption{(a) Average occupation of dot 1 as a function of the
magnetic field detuning $\delta_{2}$ when the target spin is up
(solid line) and down (dashed line) for the parameters
$J=4\gamma$, $\gamma=1.5\times10^{-6}U$, $k_{B}T=U/100$,
$\hbar\Omega=10^{-6}U$, $\hbar\omega_{0}=10^{-3}U$. (b) The same
as in (a) for $J=0.4\gamma$.} \label{spinreso}
\end{figure}

Inspection of the $\sigma_{i}^{z}$ terms in Hamiltonian
Eq.~\ref{hamiltf} shows that the interaction of the nearby spin
with the spins on dot 1 induces an effective Zeeman asymmetry
between the two dots of order $J$ that depends on the orientation
of the nearby spin. This suggests that a microwave-induced change
in the occupation of the DD could take place and reveal
information about the spin state when the dot parameters are
adjusted to the spin blockade regime.

A Zeeman asymmetry can in principle arise due to intrinsic factors
as in the case where the two coupled dots have different
$g$-factors leading to $\Delta_{1}\neq\Delta_{2}$ which could make
the spin detection difficult. In Ref.~17 we have shown how to
detect a magnetic field gradient and/or a difference in the
$g$-factors in the absence of the nearby spin that corresponds to
$J=0$. Within our scheme spin detection is efficient when the
intrinsic Zeeman asymmetry is much smaller than the spin
interaction, i.e., when $|\Delta_{1}-\Delta_{2}|\ll J$.
Nevertheless, in this work we assume for simplicity that
$\Delta_{1}=\Delta_{2}$ and thus the Zeeman asymmetry is due
entirely to the presence of the nearby spin, and further that $J$
is independent of the applied magnetic field.

\begin{figure}
\begin{centering}
\includegraphics[height=6.5cm]{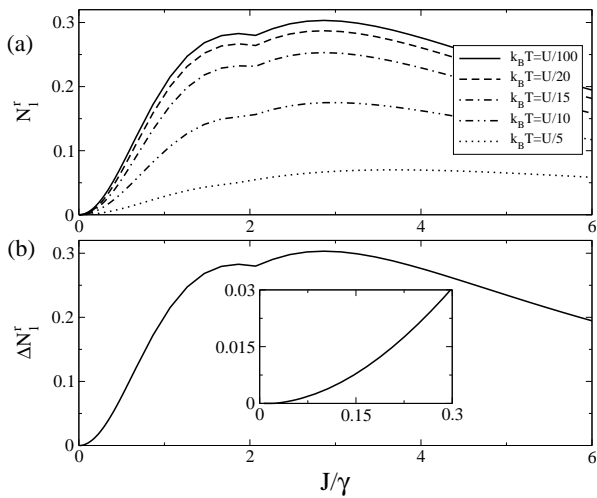}
\par\end{centering}
\caption{(a) Microwave-induced change in the occupation of dot 1
as a function of the interaction strength $J$ for different
temperatures and when the target spin is up and
$\gamma=1.5\times10^{-6}U$, $\hbar\Omega=10^{-6}U$,
$\hbar\omega_{0}=10^{-3}U$. (b) Maximum difference of the spin up
and down occupations for the same parameters as in (a) with
$k_{B}T=U/100$. The inset is an enlarged view for small $J$.}
\label{spinJ}
\end{figure}

For the numerical calculations the internal parameters of the DD
are adjusted to zero energy detuning and specifically we choose
$E(1,1)-E(0,2)=0$, with $E(n,m)$ the energy of the charge state
$(n,m)$. For a practical realization this configuration could be
achieved via adjusting the gate voltages that define the confining
potential of the double dot.\cite{koppens2} We choose for the
interdot Coulomb energy $V=U/2$ and for the on-site energies
$\varepsilon_{1}=-U/2$ and $\varepsilon_{2}=-U$. The bias voltage
is $V_{sd}=(\mu_{L}-\mu_{R})/e=U/2e$ and is applied symmetrically,
thus $\mu_{L}=U/4$ and $\mu_{R}=-U/4$, with $\mu_{L}(\mu_{R})$
being the chemical potential of the left(right) lead. When the
interdot hopping satisfies $\gamma\ll U$ and the temperature
$k_{B}T\lesssim U/80$ the current as a function of source-drain
bias is suppressed due to spin blockade and each dot contains a
single electron.

Figure~\ref{spinreso}(a) shows the average electron occupation of
dot 1 as a function of the magnetic field detuning $\delta_{2}$
for the two possible states of the target spin. We consider no
spin relaxation and therefore we set
$\Lambda_{1,j}=\Lambda_{2,j}=0$ in Eq.(\ref{bathint}). The dot
occupation exhibits resonances (peaks) due to intradot spin
rotations induced by the oscillating magnetic field and interdot
tunneling, and it is constant close to unity far from the
resonances due to spin blockade. For each spin configuration there
are in total four peaks whose positions depend on $\gamma$ and
$J$. The two outer peaks correspond to resonances R1 and R3 (see
Fig.~\ref{diagram}) and we would expect a third resonance at
$\hbar\omega_{0}=(\Delta_{1}+\Delta_{2}\pm J)/2$, corresponding to
R2, but this resonance is split by an antiresonance, resulting in
the two inner peaks. In terms of spin dependent detuning the
condition for the antiresonance is $|\delta_{1}\pm
J|=|\delta_{2}|$. This very symmetric situation together with the
microwave driving leads to the emergence of an eigenstate of the
system with purely $(1,1)$ triplet components, thus leading to
spin blockade. It can be shown from the steady-state occupations
that for this detuning the $|S_{02}\rangle=|0,\uparrow\downarrow>$
state is unoccupied and the current is suppressed. Away from the
symmetry point, when $\kappa>\gamma\hbar\Omega/J$, defining the
distance from the symmetry point as $\kappa=|\delta_{1}\pm
J|-|\delta_{2}|$, the eigenstate is no longer a pure triplet and
current can flow, resulting in the two inner peaks left and right
of the antiresonance. We can try to gain an intuitive
understanding of the outer peak positions. Naively one would
expect the outer peaks to appear when either the spin in dot 1 is
on resonance or the spin in dot 2 is on resonance. However, for
finite interdot hopping the outer peaks are somewhat shifted from
the positions that one would expect for independent spins since
intradot spin rotations take place with interdot hopping. As a
result the shift is large when $\gamma$ is large. Interdot hopping
leads to delocalisation of the electron spins via the resonant
coherent transitions
$|\uparrow,\downarrow\rangle\leftrightarrow|0,\uparrow\downarrow\rangle$
and
$|\downarrow,\uparrow\rangle\leftrightarrow|0,\uparrow\downarrow\rangle$,
with an amplitude proportional to $\gamma$, that populate the
$|0,\uparrow\downarrow\rangle$ state and thus lead to a change in
the populations. In addition, the populations of the
$|\uparrow,\downarrow\rangle$ and $|\downarrow,\uparrow\rangle$
states are unequal leading to a mixing of the (1,1) singlet and
the $S_{z}=0$ triplet that depends on the magnitude of $J$ and
$\gamma$. When $J$ is small the two outer peaks overlap with the
inner peaks as shown for example in Fig.~\ref{spinreso}(b), though
the important point is that the spin up and down signals can still
be distinguished (see also below) making the measurement feasible.
When $J=0$ resonances do not occur since the spins in the two dots
rotate at the same rate and the $|S_{02}\rangle$ state remains
unoccupied.

\begin{figure}
\begin{centering}
\includegraphics[height=15cm]{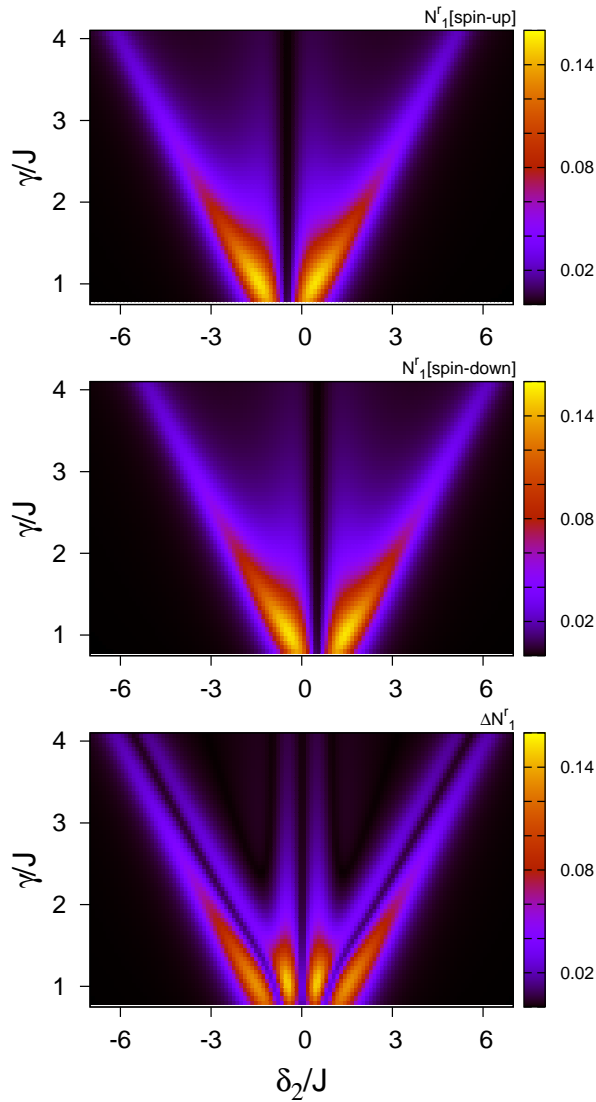}
\par\end{centering}
\caption{Relative occupation of dot 1 as a function of magnetic
field detuning $\delta_{2}$ and interdot hopping $\gamma$ for spin
up and down for the parameters $J=9\times 10^{-7}U$,
$k_{B}T=U/100$, $\hbar\Omega=10^{-6}U$,
$\hbar\omega_{0}=10^{-3}U$. The lower frame shows the absolute
difference of the spin up and down occupations.}\label{spin3D}
\end{figure}

The change in the dots' occupations is large when the Rabi
frequency, $\Omega$, is smaller than $\gamma/\hbar$ or the same
order of magnitude, a condition that allows interdot hopping while
intradot spin rotations take place. In the opposite limit spin
rotations dominate and the average dot occupation remains close to
unity in both dots. If $J$ is much smaller than $\gamma$ the
mixing of two-electron states is weak, whereas in the opposite
limit interdot transitions are suppressed and the microwaves have
no significant effect resulting in a rather small resonant change.
Therefore, probing a small $J$ needs a small interdot hopping
$\gamma$.

To quantify this effect we show in Fig.~\ref{spinJ}(a) the
relative occupation on resonance as a function of $J$ for a fixed
interdot hopping and for different temperatures when the nearby
spin is up. A spin down results in the same behaviour. The
relative occupation of dot 1, $N_{1}^{r}$, is calculated as
$N_{1}^{r}=N_{1}^{0}-N_{1}^{p}$ where $N_{1}^{p}$ is the
occupation on resonance, i.e., the value of the occupation at the
strongest peak, and $N_{1}^{0}$ is the occupation off resonance
(background occupation). The weak feature in the curves at
$J\sim2\gamma$ is due to the outer peaks (R1 and R3 regions in
Fig.~\ref{diagram}) becoming distinguishable from the inner peaks
(R2 region). For temperatures $k_{B}T\lesssim U/80$ the background
occupation is fixed, $N_{1}^{0}\sim 1$, due to spin blockade and
hence the relative occupation is essentially temperature
independent. In this regime there is to good approximation one
electron in each dot in the $|T_{\pm}\rangle$ states. With
increasing temperature spin blockade is gradually lifted and the
background occupation increases. All one- and two-electron states
acquire a finite population and even three-electron states, for
instance $(1,2)$, become occupied and have to be included in the
dynamics of the density matrix. This happens since the lengthening
tail of the Fermi-Dirac distribution of the lead electrons leads
to the opening of additional transport channels. The exact
temperature dependence of $N^{r}_{1}$ depends on various factors
such as coupling to the leads, spin relaxation rate, as well as
the applied source-drain bias. Even though this dependence may not
be monotonic in all cases for high enough temperatures
($k_{B}T\gtrsim U$) the resonances cannot be clearly resolved and
$N_{1}^{r}\sim 0$. From Fig.~\ref{spinJ}(a) we conclude that the
DD detector has a higher temperature range of operation compared
with a single dot, since the charging energy is the relevant
energy scale. A similar increase in operating temperature has been
predicted for an undriven DD read-out of a charge
qubit.\cite{gilad} In a spin read-out situation we are not only
interested in the height of the resonant peaks, but we want to
distinguish two target spin states. Thus the figure of merit for a
spin read-out has to be the maximum difference in population for
target spin up and target spin down. In Fig.~\ref{spinJ}(b) we
plot the maximum difference $\Delta N^{r}_{1}$ of the spin up and
down occupations as a function of $J$ and for a fixed interdot
hopping. For $J\gtrsim2\gamma$ the maximum difference occurs at
the inner peak of the spin up (down) occupation with
$\delta_{2}<0$ ($\delta_{2}>0$). The results indicate that a large
difference can be induced making possible the discrimination
between spin up and down states.

As shown above the achievable difference in dot occupation depends
on a range of parameters. Figure~\ref{spin3D} shows a contour plot
of the average occupation as a function of detuning and interdot
hopping for a fixed spin interaction strength.\cite{note2} The
occupation exhibits a distinct resonant pattern for both spin up
and down and further it enables the two possible outcomes to be
distinguished in a range of interdot hopping. Our calculations
confirm that this is a robust behaviour that occurs for other
values of $J$ in the range $\sim(10^{-7}-10^{-6})U$. However, as
explained above for $\gamma\gg J$ the occupation peaks decrease
and this could make the spin detection relatively difficult.

In addition to the temperature effect the background average
occupation increases due to spin relaxation and as a result the
microwave-induced resonances cannot be clearly resolved since the
relative occupation in both dots drops. Spin relaxation and
decoherence will also influence the peak height of the resonances,
since they inhibit coherent spin rotations. Spin-flip processes
which take place because of the interaction of the DD spins with
the bosonic bath described by Eq.(\ref{bathint}) allow incoherent
transitions between two-electron states, for example
$|\uparrow,\uparrow\rangle\leftrightarrow|\downarrow,\uparrow\rangle$,
$|\uparrow,\downarrow\rangle$ which in turn populate the
$|S_{02}\rangle$ state, lifting the spin blockade, and thus
increasing the background occupation. This happens even in the
absence of the nearby spin, i.e., when $J=0$ though spin blockade
can still be recovered as shown in Ref.~16 depending on the spin
relaxation rate and the coupling to the leads.

\begin{figure}
\begin{centering}
\includegraphics[height=5.5cm]{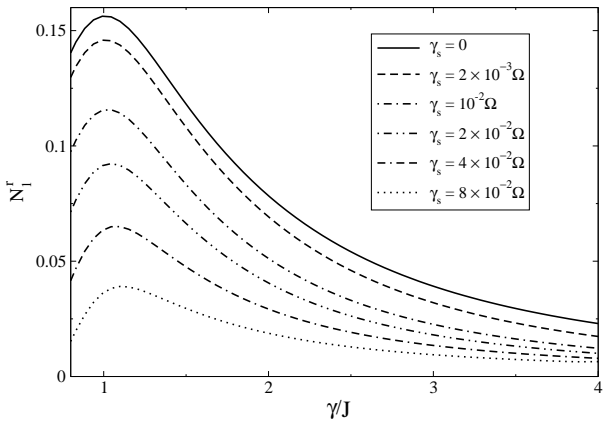}
\par\end{centering}
\caption{Relative occupation of dot 1 as a function of the
interdot hopping for different spin relaxation rates when the
target spin is up and the parameters $J=9\times 10^{-7}U$,
$k_{B}T=U/100$, $\hbar\Omega=10^{-6}U$,
$\hbar\omega_{0}=10^{-3}U$. A spin down shows the same behaviour.}
\label{spingamma}
\end{figure}

To examine the effect of spin relaxation on the driven DD spin
detector we have calculated the change in population for various
spin relaxations rates $\gamma_{s}$. Results are shown in
Fig.~\ref{spingamma} for the relative occupation of the strongest
(outer) peaks as a function of hopping when the nearby spin is up
(the same behaviour results for spin down). We have taken
$\gamma_{s}=\pi|\Lambda|^{2}D(\delta\epsilon)[2n(\delta\epsilon,T)-1]/\hbar$,
with the Bose function
$n(\delta\epsilon,T)=[\exp(\delta\epsilon/k_{B}T)-1]^{-1}$, and
$\delta\epsilon=\hbar\omega_{0}$. $D$ is the density of states for
the bosonic bath that is taken constant and also
$|\Lambda|=|\Lambda_{1,j}|=|\Lambda_{2,j}|$ in Eq.(\ref{bathint}).
This expression for $\gamma_{s}$ can be derived by assuming a
single spin with Zeeman splitting $\delta\epsilon$ coupled to a
bosonic bath at temperature $T$ which we assume to be the same as
the temperature in the leads. We focus on the most interesting
experimental regime in which the spin relaxation rate is much
smaller than the tunneling rate through the double dot and the
system weakly deviates from the spin-blockade regime. As seen in
Fig.~\ref{spingamma} the effect of spin relaxation becomes
important as $\gamma_{s}$ increases which in turn leads to a
decrease in the relative occupation. The sensitivity of the spin
detector is limited by the minimum detectable change in the
occupations. The optimum resolution can be achieved when the
driving is efficient. This happens when the Rabi frequency, which
is controlled by the intensity of the oscillating field, is larger
than the spin relaxation rate of the dot spins and the target
spin, as well as the tunneling rate through the DD.

\section{\label{sec:Conclusions}Conclusions}

In summary, we have suggested an electrical scheme to probe a
single spin which makes use of two serially tunnel-coupled quantum
dots connected to metallic leads. The spin is located at some
distance from the dots, which has to be smaller that the typical
interdot separation, and the total system is in a static magnetic
field under the application of a microwave magnetic field. The
spin interacts with the spins on the dots and this interaction
results in an effective Zeeman splitting that is different in the
two dots. Due to an electron spin resonance effect the electron
occupations of the dots exhibit resonances which reveal
information about the state of the nearby spin. In particular, the
ac-driven DD spin detector provides an explicit signal in the
induced occupations for both spin orientations and enables the
spin state to be probed noninvasively in a single shot provided
that the target spin has a different $g$-factor from the DD
system, a condition that is typically satisfied.

We identified a range of parameters for which the system can
operate and analysed how we can tune its sensitivity with the
interdot hopping and intensity of the microwave field that defines
the Rabi frequency for spin rotations. The operation of the DD
detector depends on a lifting of a Pauli spin blockade and
therefore it can operate at much higher temperatures than the
single dot which is limited to temperatures comparable to the
Zeeman energy.\cite{wabnig1} For instance for a charging energy of
10 meV and interaction strength in the range $\sim$5 MHz the
resonances survive up to temperatures of a few tens of Kelvin. To
achieve a similar operating temperature with a single dot,
magnetic fields of a few tens of Tesla and correspondingly
microwave frequencies of several hundreds of GHz would be
necessary, conditions which are available only in specialised
laboratories. The sensitivity of the detector is limited by
internal spin relaxation which essentially leads to a small change
in the occupations and as a consequence the resonances cannot be
clearly resolved. For an efficient read-out the tunnelling rates
from dots to leads and the microwave-induced Rabi frequency have
to be larger than all the relevant spin relaxation rates.

Finally, the change in population effected by the microwave field
has to be seen in relation to the relaxation time of the target
spin. On one hand, the spin read-out has to be finished within the
relaxation time of the target spin, otherwise random spin flips
will obscure the result. On the other hand, for a given change in
population, $\Delta N_{i},$ of dot $i=1,2$ a certain number of
electrons, $n$, has to pass through the double dot and be counted
by the charge detector. To achieve reliable statistics the
requirement $n>1/\Delta N_{i}^{2}$ has to be fulfilled, since then
fluctuations in the average number are smaller than the change in
population that we want to distinguish. At a given tunneling rate
$\Gamma$ through the device this determines the minimum time of
the measurement $T_{m}=n/\Gamma=1/(\Delta N_{i}^{2}\Gamma)<T_{1}$,
which must be smaller than the spin relaxation time of the target
spin. We arrive therefore at a minimum tunneling rate through the
dot: $\Gamma>1/(\Delta N_{i}^{2}T_{1})$. However, we are not free
to increase the tunneling rate arbitrarily; once the tunneling
rate approaches the Rabi frequency the resonance peaks in the
population disappear. By demanding $10\Gamma<\Omega$ we obtain the
minimum resolvable change in population as $\Delta
N_{i}=\sqrt{10/(T_{1}\Omega)}$.

We conclude by estimating the feasibility of measuring the state
of a molecular spin system, for example $\textnormal{Sc@C}_{82}$,
which can be coupled with a carbon nanotube double quantum
dot.\cite{buitelaar} Such carbon based systems are promising
candidates for quantum information processing and around several
Kelvin have a $T_{1}\approx1$ s (Refs.~20, and 21). Then with a
Rabi frequency of $10$~MHz we arrive at a minimum resolvable
change in population of $\Delta N_{i}=0.001$. An interdot hopping
of $\gamma=10$~MHz and a spin-spin interaction of $J\approx5$~MHz
would lead to $\Delta N_{i}\approx0.07$ in the case of no spin
decoherence. With a spin decoherence rate of $\gamma_{s}=1$~MHz
this reduces to $\Delta N_{i}\approx0.007$, indicating that a
single spin read-out with realistic parameters at liquid helium
temperatures is feasible.

\section*{ACKNOWLEDGEMENTS}

We would like to thank Mark Buitelaar for discussions. The work is
part of the UK QIP IRC (GR/S82176/01). JW thanks the Wenner-Gren
Foundations for financial support. BWL thanks the Royal Society
for a University Research Fellowship. JHJ acknowledges support
from the UK MoD. GADB is supported by an EPSRC Professorial
Research Fellowship (GR/S15808/01).


\begin{thebibliography}{10}
\bibitem{note1}
We are interested in static magnetic fields which
produce a Zeeman splitting on the order of $~\hbar\omega_{0}$ and
small interdot hoppings and spin interaction strengths. In this
regime the $|T_{-}\rangle(|T_{+}\rangle)$ is the lowest (highest)
two-electron state.
\bibitem{note2}
The interdot hopping cannot be made arbitrarily
small, since this will invalidate the Born approximation.
\end{thebibliography}
\end{document}